\documentclass{emulateapj}
\newcommand{\kms}{km~s$^{-1}$}
\newcommand{\nick}{$^{56}$Ni}
\newcommand{\gcm}{g~cm$^{-3}$}
\shorttitle{Signature of Electron Capture in SN~2003du}
\shortauthors{H\"oflich et al.}
\begin{document}
\title{Signature of Electron Capture in Iron-Rich Ejecta of SN~2003du \footnotemark[1]}
\author{
Peter H\"oflich\altaffilmark{2},
Christopher L. Gerardy\altaffilmark{2,3},
Ken'ichi Nomoto\altaffilmark{4}, 
Kentaro Motohara\altaffilmark{5},
Robert A. Fesen\altaffilmark{6},
Keiichi Maeda \altaffilmark{7},
Takuya Ohkubo\altaffilmark{4},
Nozomu Tominaga\altaffilmark{4},
}
\footnotetext[1]{Based on data collected at Subaru Telescope, which is operated by the
National Astronomical Observatory of Japan, and at McDonald Observatory
of the University of Texas at Austin.}
\altaffiltext{2}{McDonald Observatory, University of Texas at Austin, Austin, TX 78712}
\altaffiltext{3}{W. J. McDonald Postdoctoral Fellow}
\altaffiltext{4}{Department of Astronomy, University of Tokyo, Bunkyo-ku, Tokyo 113-0033, Japan}
\altaffiltext{5}{Institute of Astronomy, University of Tokyo, Mitaka, Tokyo, Japan}
\altaffiltext{6}{Department of Physics \& Astronomy, Dartmouth College, 6127 Wilder Laboratory, Hanover, NH 03755}
\altaffiltext{7}{Department of Earth Science and Astronomy, College of Arts and Sciences, University
 of Tokyo, Meguro-ku, Tokyo 153-8902, Japan}

\begin{abstract}

Late-time near-infrared and optical spectra of the normal-bright Type Ia supernova 2003du  
about 300 days after the explosion are presented. 
At this late epoch, the emission profiles of well isolated [\ion{Fe}{2}] lines (in particular that of 
the strong 1.644 \micron\ feature) trace out the global kinematic distribution of radioactive material in
the expanding supernova ejecta.  In SN~2003du, the 1.644 \micron\ [\ion{Fe}{2}] line seems to show a flat-topped,
profile, indicative of a thick but hollow-centered expanding shell, rather than a strongly-peaked
profile that would be expected from a ``center-filled'' distribution.
Based on detailed models for exploding Chandrasekhar mass white dwarfs,
we show that the feature is consistent with spherical explosion models. Our model predicts
 central region of non-radioactive electron capture elements
up to 2500--3000 \kms\ as a consequence of burning under high density, and
 an extended region of radioactive \nick\
up to 9,000--10,000 \kms.  Furthermore our analysis indicates that the 1.644 \micron\ [\ion{Fe}{2}]
line profile is not consistent with strong mixing between the regions of electron-capture
isotopes and the \nick\ layers as is predicted by detailed 3D models for nuclear deflagration fronts. 
We discuss the possibility that the flat-topped profile could be produced as a result of an infrared
catastrophe and conclude that such an explanation is unlikely.
We discuss the limitations of our analysis and
place our results into context by comparison with constraints on the distribution of
radioactive \nick\ in  other SNe~Ia and briefly discuss the potential implications of our result
for the use of SNe~Ia as cosmological standard candles.

\end{abstract}
 
\keywords{supernovae -  nuclear burning fronts - progenitor systems}

\section{Introduction}
 
There is general agreement that Type Ia Supernovae (SNe~Ia) are the result of a thermonuclear explosion
of a degenerate C/O white dwarf (WD) most likely close to the Chandrasekhar limit for the majority of events,
because these allow to reproduce  optical/infrared light curves (LC) and  spectra of SNe~Ia reasonably well.
 However, despite the success of Chandrasekhar mass models,
 questions remain concerning the detailed physics of the burning
front, the ignition process, the role of pre-conditioning and progenitor
of the WD prior to the explosive phase.
For recent reviews see \citet{branch99}, \citet{hetal03}, \citet{nomoto03}.

The distribution of  the iron-group elements  is key to answer these questions.
It has long been recognized that late-time spectra of SNe~Ia, dominated by forbidden lines of Fe-peak 
elements, are an excellent probe of  the amount and distribution of radioactive \nick\ \citep{a80a,a80b,m80}.
After about day 200, the energy input is dominated by local energy deposition from positrons
and the envelope is almost transparent, allowing for the direct observation of the entire distribution of 
radioactive matter, and making the result rather insensitive to radiation transport effects.
However, the total amount of Ni is still rather uncertain because the emission
depends sensitively on the redistribution of energy within an ion (see also IR catastrophe; 
\citealt{a80a,fransson96}), and on the ionization mechanism  \citep[e.g.][]{s92,l97,b97}.

In principle, line profiles are less sensitive to these effects
since they depend not on the absolute amount of excitation at a certain time but rather on the relative 
change of the conditions within the envelope. Unfortunately the spectral features in the optical 
are produced by a large number of overlapping lines, making the intrinsic emission
profiles difficult to untangle.  The near-IR contains much more isolated lines, potentially offering the
opportunity to look for the remnants of high-velocity plumes in the outer regions of the \nick-rich
gas and to look for kinematic offsets of the \nick\ distribution from off-center ignition.  
However, infrared spectroscopic observations at such late epoch are difficult to obtain and to-date, very 
late-time ($> 200$d) IR spectra have only been published for the unusual supernova SN~1991T 
\citet{p93,s92,b97}.

Here we present a study of late-time spectra of the normal bright Type~Ia SN 2003du, concentrating on 
NIR spectroscopic observations with the 8.2m Subaru telescope.  In \S~2, we discuss the observations and
data reduction. In \S~3, we describe our models used for the analysis of observations and its implications
which is presented in \S~4.
In \S~5, we discuss the results with an eye toward the broader context. In \S~6, we sum up the results and
and comment on the limits of our study and future perspectives.

\section{Observations and Data Reduction}
SN~2003du was discovered in UGC~9391 by LOTOSS \citep{schwartz03} on 22 April 
2003 (UT) at about $15.9^m$. It was classified as Type~Ia by \citet{kotak03} 
on 24 April 2003 who noticed that it resembled SN~2002bo about two weeks before maximum light.
Early-epoch optical spectroscopy revealed a high-velocity component of the
\ion{Ca}{2} infrared triplet feature, similar to that seen in SN~2001el, which
may be the signature of interaction with circumstellar material, but otherwise
the optical evolution appeared normal \citep{gerardy04}.

\begin{figure*}
\includegraphics[width=14.7cm,angle=360,clip=true, trim=60bp 0bp 0bp 130bp]{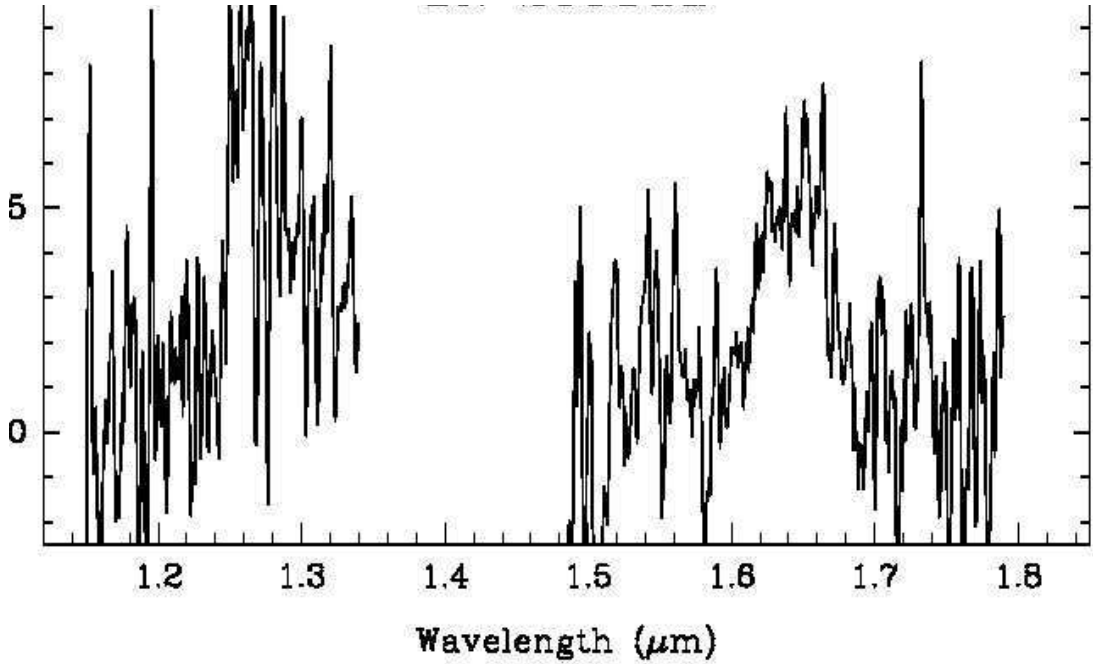}
\caption{Near IR spectrum of SN2003du on February 27th, 2004, obtained at the Subaru telescope with the OH-Suppressor
spectrograph (OHS).  The 2-D data are shown after the data have been pair-subtracted and stacked, but prior to 
extraction and correction for instrumental response. Note that the S/N is highly wavelength dependent due to the
influence of night-sky emission features even with the OH-suppressor. }
\label{IR_spectrum}
\end{figure*}

\begin{figure*}
 \includegraphics[width=10.7cm,angle=270,clip=true, trim=0bp 0bp 0bp 0bp]{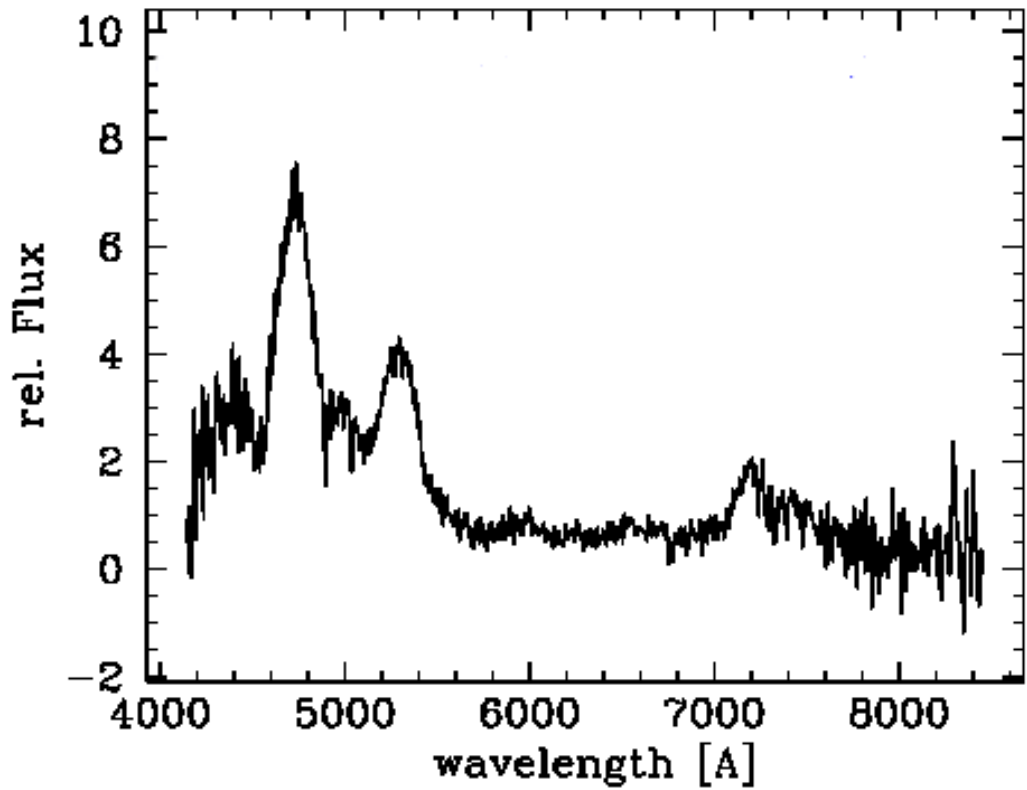}
\caption{Optical spectrum of SN2003du on March 27, 2004, normalized at 7000 \AA.}
\label{spectra}
\end{figure*}

\subsection{NIR Spectroscopy}
A near-infrared spectrum (1.1--1.8 \micron\, R$\approx 400$, Fig.~\ref{IR_spectrum})
of SN~2003du, was obtained on 27 Feb 2004 (296 days after $V_{max}$) using the
OH-airglow Suppressor \citep[OHS][]{iwamuro01} and the Cooled Infrared 
Spectrograph and Camera for OHS \citep[CISCO][]{motohara02} on the Subaru 
8.2~m telescope.  The observations consisted of two 2000~s exposures, using a 
0\farcs5 slit, with the object dithered 10\arcsec\ along the slit between
exposures.  A second, lower-S/N spectrum, totaling 1500~s on-target 
integration time, was obtained on 2 Apr 2004, (331 days after $V_{max}$). 

Data reduction was performed using standard IRAF routines \citep{tody86}. The two exposures
were differenced for first-order sky subtraction, then the ``negative'' 
spectrum was inverted, shifted, and combined with the 
``positive'' spectrum.  Cosmic rays, hot pixels and other detector defects
were removed using the LA-Cosmic package \citep{vandokkum01}.
Since the 0\farcs105 pixels oversample the spectrum,
the combined data were boxcar filtered 3x3 before 1-D spectra were extracted.
During the extraction, a second background subtraction was performed using the
average background level from the regions adjacent to the extraction aperture.
1-D spectra were wavelength calibrated using the standard OHS/CISCO wavelength
mapping\footnote{For details, refer to the Subaru OHS/CISCO website:
http://www.naoj.org/Observing/Instruments/OHS/index.html},
shifted to the measured slit position during the observations.  The combined
atmospheric and instrumental transmission function was removed from the data
by following the target observations with observations of an A2 star (SAO029603) at similar
airmass, and matching the observed spectrum of the telluric star with
stellar atmosphere models of Kurucz (1994).

The S/N of the resulting spectrum is highly wavelength dependent.  The data are
completely sky-noise dominated so the S/N is essentially independent of the 
signal strength from the supernova, but is instead strongly dependent on the level of the
night-sky emission at a given wavelength.  Where strong night-sky emission remains
after passing through the OH-suppressor, a narrow region of low-S/N is created.
This effect can be seen very clearly in the 2-D pre-extraction image shown in 
Figure~\ref{IR_spectrum}, where the bright O$_2$ band near 1.3 \micron\ is not
suppressed by the OHS and creates a band of very low S/N in the resulting data.


The reduced NIR spectrum of SN~2003du on 27 Feb is shown in Figure~\ref{IR_spectrum}, 
with the wavelengths presented in the rest-frame of the host galaxy ($cz=1914$ \kms; \citealt{schneider92}).
The spectrum exhibits broad emission from the [\ion{Fe}{2}] 1.257~\micron\ (a$^6$D$_{9/2}$--a$^4$D$_{7/2}$)
and 1.644~\micron\ (a$^4$F$_{9/2}$--a$^4$D$_{7/2}$) lines which typically dominate nebular Fe spectra in 
the NIR \citep[e.g.][]{ns80,omd89,omd90}.
Surprisingly, the 1.644 \micron\ line appears to show a flat-topped
line profile suggestive of an extended envelope with a central hole in the emissivity. In contrast,
a filled envelope would produce a strongly peaked line profile.  
The S/N of the 1.257 \micron\
feature is too low to make any strong statements about the shape of the line profile,
but it appears to be consistent with the boxy shape of the 1.644 \micron\ feature.
Similarly, the low S/N spectrum on 2 Apr is consistent with no significant evolution from
the 27 Feb spectrum.
Understanding the nature and implications of this apparent flat-topped emission profile 
is the subject of our analysis. 

\subsection{Optical Spectroscopy}
In addition to the NIR observations, optical spectra (4200--8000 \AA, R$\approx 
450$) of SN~2003du were obtained
on 25 Mar 2004 (323 days after $V_{max}$), using the Imaging Grism Instrument
on the 2.7~m Harlan J. Smith Telescope at McDonald Observatory.  Observations
of SN~2003du consisted of four 3600~s exposures using a 2\arcsec\ E--W oriented
slit.  Wavelength calibration was achieved by observing Cd, Ar, and Ne arc
lamps and the spectra were flux calibrated using the \citet{massey88} standards
G191B2B, Hiltner 600, Feige 34, and PG 0823+546.

The reduced optical spectrum is shown in Figure~\ref{spectra}.  The spectrum is
typical of late-time optical spectra of SNe~Ia \citep[e.g.][]{b97}, showing
a strong blend of forbidden lines from Fe-peak elements in the 4000--5000 \AA\ 
region.  The feature near 7300 \AA\ is the comparatively isolated [\ion{Fe}{2}]
7155 \AA\ line, blended with [\ion{Ca}{2}] 7291 \& 7324 \AA\ emission and other
weak [\ion{Fe}{2}] lines at 7142, 7438 \& 7648 \AA .

\section{Model Construction}

Our quantitative comparison is based on the delayed-detonation model {\sl 5p0z22.23} which
reproduces the  light-curve and spectra of a normal-bright SN~Ia like SN2003du reasonable well
 \citep{h02,gerardy04}.
In delayed-detonation models (DD), the white dwarf progenitor undergoes supersonic detonation after a period
of pre-expansion, perhaps due to a phase of subsonic deflagration  \citep{k91,yamaoka92,ww94}.
We choose a delayed detonation model because 
this class of model have been found to reproduce the majority of the observed
optical/infrared light curves (LC) and  spectral features of SNe~Ia including the brightness
decline relation \citep{h95,fisher95,h96,hk96,wheeler98,lentz01,h02}.
As a reference, we also examine the profile in the context of the well-known deflagration model W7 \citep{nomoto84}.

 Our results are more generally applicable to all Chandrasekhar-mass models since the structure
of the WD, the explosion energy, and the light-curves are mainly determined by nuclear physics 
rather than the details of the nuclear burning (``stellar amnesia''; \citealt{hetal03}). 
 We use these specific models as  testbeds to interpret the late-time observations.

\subsection{Open Questions of the Explosion Physics}
Despite the successes of delayed detonation models (and W7), questions remain concerning the detailed physics of the burning
front \citep{liv93,k95,r99,k01,r02,g02}, the ignition process, and the role of pre-conditioning
of the WD prior to the explosive phase \citep{hwt98,l00,hs02}. Overall, the
propagation of the detonation front is well understood but the description of the deflagration front
and, in the DD scenario,  the mechanism for a deflagration to detonation transition (DDT) are persistent unsolved problems as is
the location of the DDT which if off-center may result in off-center distributions of \nick.

On a microscopic scale, a deflagration propagates due to heat conduction by electrons. Though the laminar 
flame speed in SNe~Ia is well known, the
front has been found to be Rayleigh-Taylor (RT) unstable increasing
the effective speed of the burning front \citep{nomoto82}.
Current hydrodynamic calculations, starting from central ignitions in static WDs, all  
show that RT instabilities govern the morphology of the burning front in the regime of linear instabilities,
i.e.\ as long as  perturbations remain small. During
the first seconds after the runaway, the increase of the flame surface due to RT instabilities
remains small and the effective burning speed is rather small resulting in the rise of large
plumes of burned matter through the WD. 

\citet{n97} studied the effect of off-center ignition and demonstrated that
multiple-spot ignitions will alter the early propagation of the flame.
In
significantly off-center ignitions, the buildup time of RT-instabilities is shorter corresponding to the
larger gravitational acceleration, producing off-center distributions of iron-group elements,
although the overall picture of large, rising plumes still remains unchanged. As a consequence (and
in contrast to spherical explosion models), we expect mixing of electron-capture products, formed by 
high-density burning, with the radioactive \nick-rich layers from lower-density burning.
Note, however, that in all these calculations the constraint of starting from a static WD is not trivial,
as the morphology (and, consequently, the effective burning speed) of blobs 
depends on small scale motions of the background produced
prior to the explosive burning phase \citep{hs02} particularly in the earliest burning stage.

Probing the properties of the deflagration burning front directly with observations has proven difficult
because the signatures of the deflagration are mostly wiped out during the detonation phase in normal-bright
SNe~Ia \citep{liv99,g03}.
 This provides the justification for our approach and the explanation why we can use a spherical DD model
as baseline.

\subsection{Model Construction}

\begin{figure*}
\includegraphics[width=6.5cm,angle=270,clip=true]{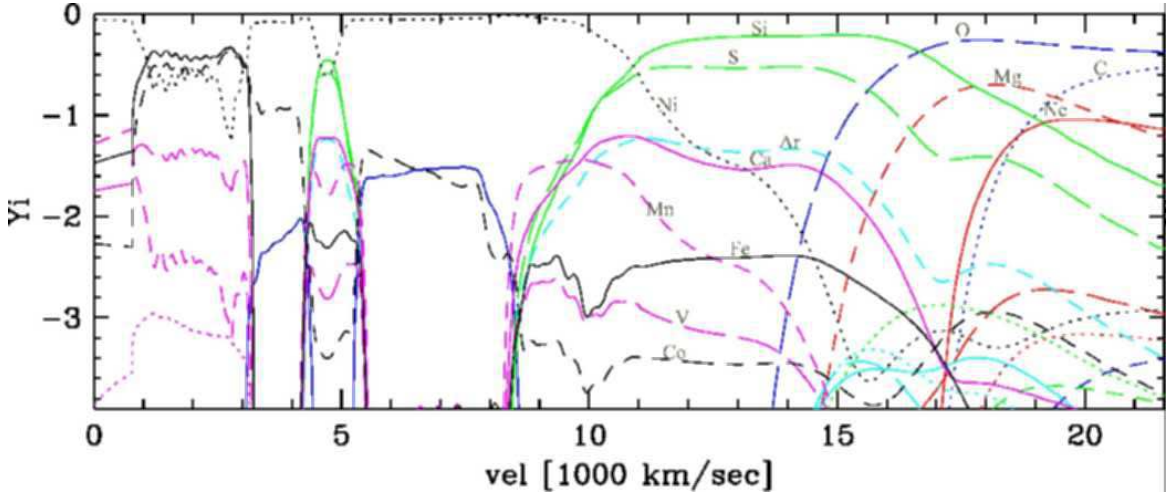}
\caption{Chemical structure of the delayed detonation model. 
The abundances of stable isotopes and $^{56}$Ni are given as a
function of the expansion velocity. The curves with the highest  abundance close to the
center are $^{54}$Fe, $^{58}$Ni and $^{56}$Fe. The calculation is based on a delayed detonation
model with central and transition densities of $2 \times 10^9$ g~cm$^{-3}$ and 
$2.5 \times 10^7$ g~cm$^{-3}$, respectively.
}
\label{struct}
\end{figure*}

 For this study, the  delayed detonation model {\sl 5p0z22.23}
(Fig.~\ref{struct}) was chosen. We use also the deflagration model
W7 as additional reference.

We consider the explosion of a Chandrasekhar mass white 
dwarf which originates from a star with a main sequence mass of 5 $M_\odot$ and solar composition.
At the time of the explosion, the central density is $2 \times 10^9$ \gcm. The nuclear
burning starts as a deflagration with a parameterized description of rate of burning based on 
3-D models by \citet{k01}. When the density reaches $\rho_{tr}=2.5 \times 10^7$ \gcm, the 
detonation is triggered. The final density and chemical structures
are given in Figure~\ref{struct}. 

The pre-expansion of the WD is determined only by the amount of burning during the
deflagration phase and not the detailed properties of the deflagration front. In 1-D spherical 
models $\rho_{tr}$ is thus a 
convenient parameterization of the amount of burning during the deflagration. After the detonation 
is triggered, the burning products depend mainly on the density under which burning occurs. Because
the detonation wave propagates rapidly through the WD at slightly above the sound speed, the
outcome is determined primarily by the density structure of the (puffed up) WD at the time of the DDT. 
As a result, the final density and chemical profile is rather generic for all DD models. 


Our analysis of the line profiles is based on radiation transport models using our Monte-Carlo scheme 
implemented in HYDRA including detailed $\gamma $-ray transport and energy deposition of fast electrons (see
H\"oflich, 2003ab, and references therein).
 Forbidden lines are included based on Kurucz's data base (1993) and
\citet{ns88a,ns88b}, and supplemented with line lists given in \citet{l97,b97}.
 Guided by detailed calculations \citep{ruiz95, l97},
 we assume that the energy input is given by the energy deposition in form of $\gamma $-rays and positrons
based on our detailed Monte Carlo calculations, and that the distribution of the dominant ion is given
by the abundance profile in the envelope which allows us to investigate line profiles.
 We do not calculate the ionization structure because its details depend sensitively on the assumptions 
and the processes. The total amount of Fe~II varies and, as a consequence, the absolute fluxes are uncertain.
For the Fe~II distribution, the main difference between models in literature are related
to the central region which may or may not show Fe~II as the dominant ionization stage.
 Here, we use the observations to constrain the conditions at the center and, thus, have
to use our approximation.

 Our approach is well supported by the current literature.
The most detailed calculations for the ionization structures near day 300
have been published for W7, a model with a chemical structure in the inner regions similar
to our delayed detonation model \citep[hereafter RL95 and LJS97, respectively]{ruiz95,l97}. Generally, 
the results of R95 and LJS97 roughly agree, namely Fe~II is the dominant ionization stage throughout the
$^{56}$Ni region with little variation in  the ionization fraction.
 Small differences, namely an increase of the Fe~III with radius,
 are present at the outer, high velocity layers which contribute to the
line wings. However,  significant qualitative difference can be seen close to the center
where R95 find that iron group elements are mostly neutral and where LJS97 find that \ion{Fe}{2} dominates
\ion{Fe}{1}. 
At an epoch of around 300 days, energy deposition and ionization by  $\gamma $-rays
is of little importance but collisions with fast electrons are the dominant source of ionization
(LJS97).  These fast electrons are produced by the energy deposition  by positrons which are produced
as a consequence of the $\beta ^+$ decay of $^{56}Co$. However, neither LJS97 nor R95 considered the effect
of  magnetic fields.  Whether the fast leptons can leak into the core depends on the physical processes considered and
on the magnetic field or, more precisely, on the Gyro-radius $R$  which is given by
$$ R = {\sqrt{2 ~m_e ~E_e} c \over e B_\bot } \backsimeq 3.4 \times 10^{3} {\sqrt{E_e}\over B_\bot} [{\rm cm}] $$
with $m_e$, $E_e$ and $e$ being the mass, energy [in MeV] and charge of the fast electron, respectively,
and $B_\bot$ the orthogonal magnetic field. Typical progenitor WD magnetic fields have been estimated to be
$\sim 10^{5}$ -- $10^{9}$ Gauss \citep{leib95}. Assuming that the flux is frozen in, the magnetic
field evolves as $B_r=B_r(t=0) \times (r_o/r)^2$. Thus by day 300, the magnetic field is about
$\sim 10^{-11}$ -- $10^{-7}$ gauss, implying that the Gyro radius is about 2\% or less of the distance to 
the inner edge of the \nick\ distribution (see Fig. \ref{struct}). Even a small $B_\bot$ would keep the
electrons local.
For comparing the line profiles as a baseline model, we assumed that the local energy input
is given by the \nick\ distribution and that the emissivity is given by the abundance distribution.

\section{Data Analysis}

\subsection{Delayed detonation scenario}
We will to concentrate on the spectral emission profile of the [FeII] at 1.644~$\mu m$ which is essentially un-blended at late times.
This feature is surrounded by weaker lines from the same multiplet, but in nebular conditions, these
lines emit at $\sim$1/3 or less of the 1.644 \micron\ feature \citep[e.g.][]{ns80,omd89,omd90}.  As such they 
may contribute some to the faint wings, but will not affect the bulk of the line profile. 

\begin{figure*}
 \includegraphics[width=12.7cm,angle=270,clip=true]{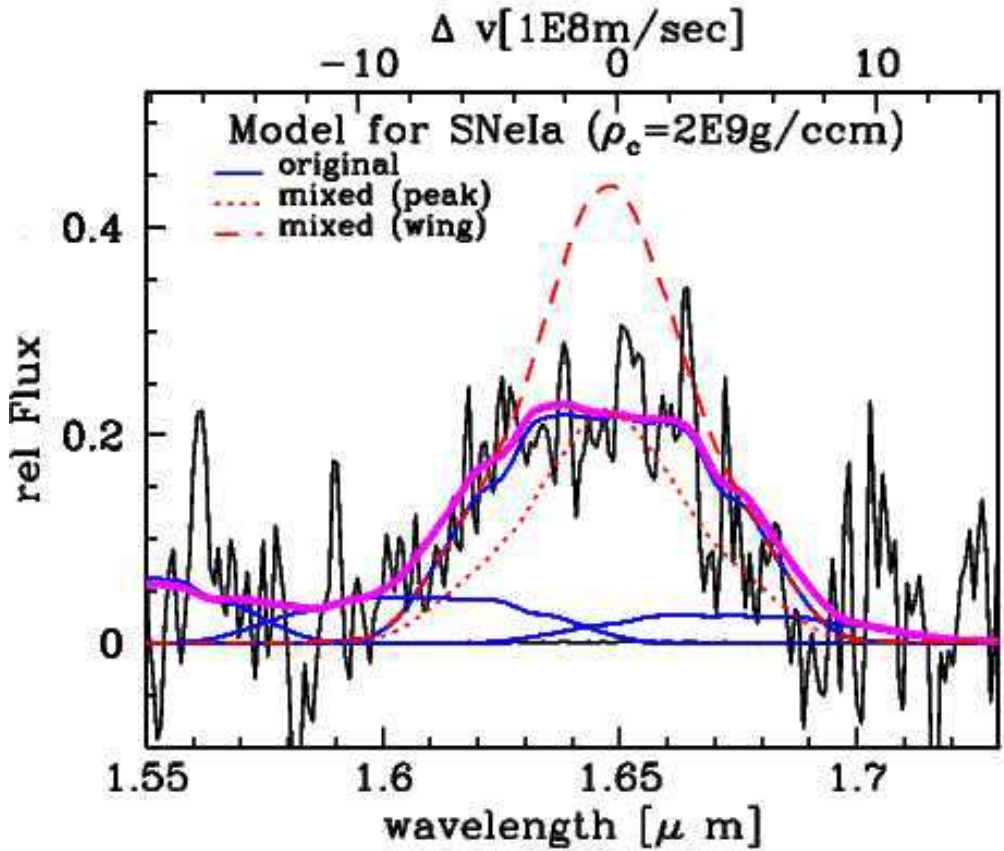}
\caption{NIR spectrum of SN 2003du on Febr. 27, 2004, in comparison with
the theoretical line profiles (solid thick line). In addition, the individual
components of  the forbidden [Fe~II] transition at 1.644 $\mu m$ are given for the original
delayed detonation model (solid) and mixed chemistry (light) normalized to the maximum
line flux (dotted) and the wings (dashed), respectively.
 In addition,  weaker individual contributions 
 are given for weaker [\ion{Fe}{2}] lines (1.599, 1.664 \& 1.677 \AA).
 The theoretical spectrum has
been shifted by the peculiar velocity of the host-galaxy ($+1932$\kms) and a residual velocity
of $-500$\kms. 
}
\label{IR_model}
\end{figure*}

\begin{figure*}
\includegraphics[width=10.7cm,angle=270,clip=true,trim=0bp 0bp 0bp 00bp]{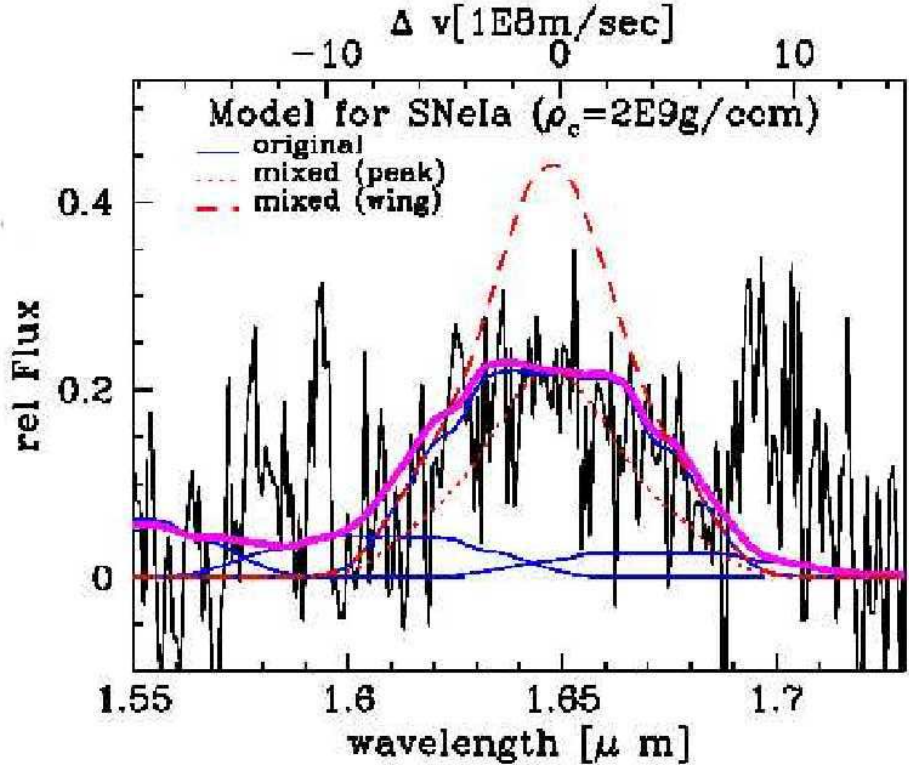}
\caption{NIR spectrum of SN 2003du on April 2nd, 2004, with 1/3 of the integration time of the 27 Feb.\ spectrum. 
The line is present and the profile is consistent with little evolution, still seeming to show a flattened profile.
However, the S/N of the spectrum is such that it cannot put strong constraints on the shape of the line profile.
The comparison line profiles are the same as for Fig.~\ref{IR_model}.
}
\label{IR_model2}

\end{figure*}

The comparison between the observation on 27 Feb and the models is given in Fig.~\ref{IR_model}.
The flat-topped profile can be well reproduced by the theoretical line profile of
our (unmixed) reference model (solid/blue line). In this model, the flat top is produced by the lack of 
radioactive matter (i.e.\ \nick/Co) close the center as a consequence of high-density burning up to an
expansion velocity of $\sim 3000$ \kms, resulting in non-radioactive electron-capture products 
(see Fig.~\ref{struct} and  also \citealt{bra00}). The extended wings of the feature are produced by the
\nick/Co region which extends from $\sim 3000$~\kms\ to $\sim 9,000$--10,000~\kms. The slight asymmetry 
in the theoretical spectrum is caused by optical depth effects due to Thomson, bound-free and free-free opacities
 and is consistent with the observed profile,
although the S/N is not nearly sufficient to actually detect such a feature.
In comparing with the observations, we shifted the model line profile 500~\kms\ to the blue. Such a shift might potentially
be a signature of an off-center \nick\ distribution, but the shift is too small ($\approx 1$ spectral
resolution element) to make any significant claim, especially given the S/N of the observed line profile.  
Due to the shorter integration time of the spectrum obtained on April 2nd, the level of noise is somewhat 
higher (Fig.~\ref{IR_model2}). The IR feature at 1.644 $\mu$m   is still present in this spectrum
at roughly the same flux level, and the profile is consistent with that seen in the earlier spectrum.

\begin{figure*}
 \includegraphics[width=10.7cm,angle=270,clip=true]{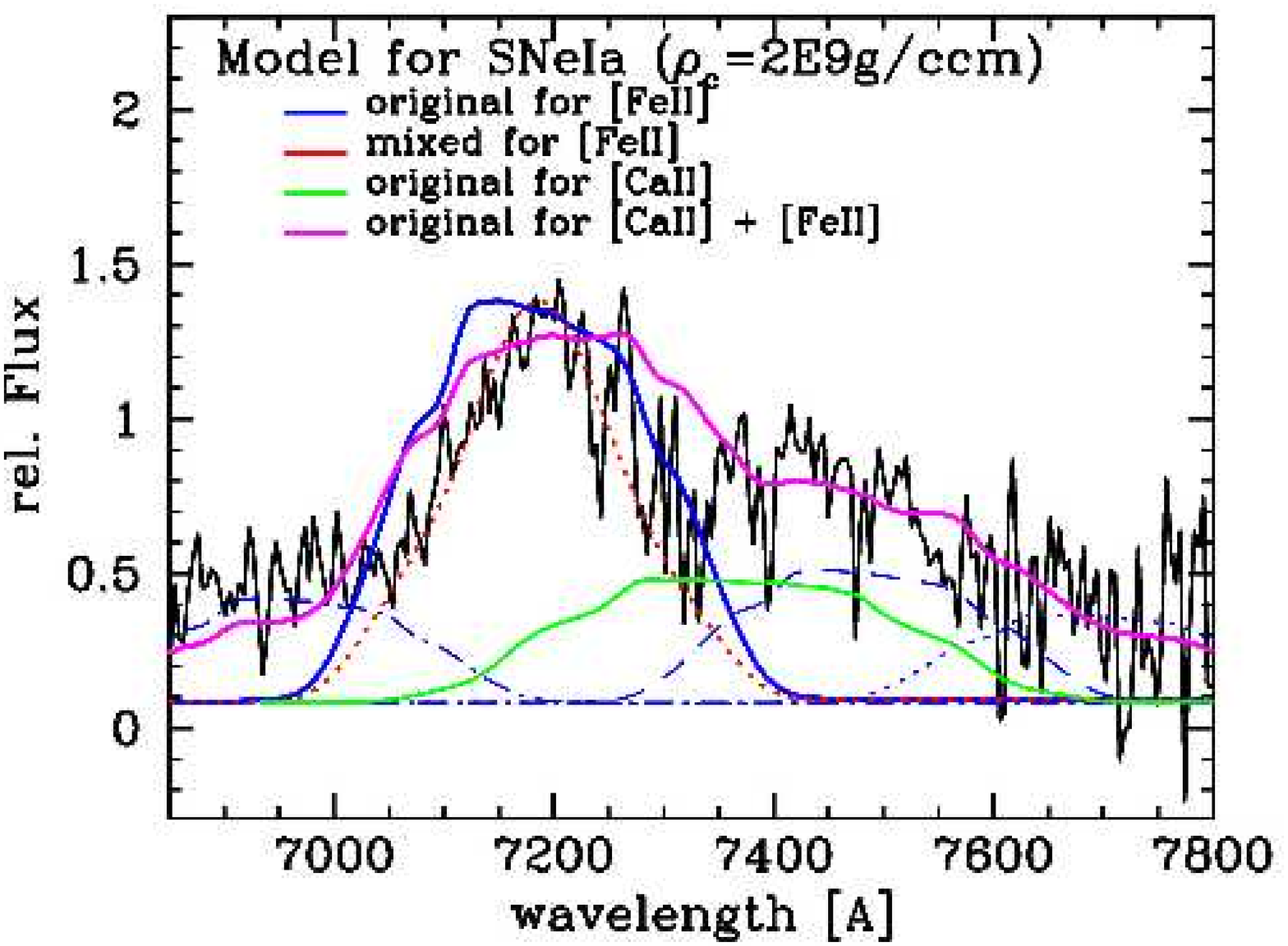}
\caption{Comparison of the observed and theoretical emission profiles (solid, dark grey) for the region between 6800 and 7790 \AA.
 Fitting the line fluxes, the  black and light-grey lines (solid) present fits of
 the [\ion{Fe}{2}](7155 \& 7172\AA) and [\ion{Ca}{2}](7291 \& 7324\AA), respectively. Contributions
of [\ion{Fe}{2}] lines are given for other weaker transitions of [\ion{Fe}{2}] (dotted and dashed-dotted).
 The dotted line shows the  [\ion{Fe}{2}](7155 \& 7172\AA) for the mixed model.
}
\label{model_optical}
\end{figure*}

In light of Rayleigh-Taylor instabilities predicted by 3-D
deflagration models, the central concentration of electron capture elements in our 1-D model
might be regarded as somewhat artificial. 3-D deflagration models produce a
mixture of radioactive and non-radioactive isotopes within the layers.  To simulate this, we also calculated
the results from a model with mixing of the inner layers below 8000 \kms. Similarly, a mixed model will also
simulate the effect of penetration of fast electrons into the core in absence of magnetic fields.

In Figure \ref{IR_model}, the resulting line profile is shown in red.
We show the profile with two normalizations, one
normalized to match the observed maximum flux and one adjusted to match the line wings.
Although the S/N of the features is low, the observations do not appear to be a good match to either normalization 
of the mixed profile.  The profile matched to the central emission is too narrow, consistently under-predicting
the flux in the outer parts of the profile, while the fit to the wings predicts significantly more flux in the 
center of the line profile than is observed.  Since the S/N of the spectrum is admittedly low, it is difficult to
absolutely rule out a more centrally peaked line profile, but the unmixed model certainly seems to be a much better
 fit to the data.  Although the low-S/N line
profile does not exclude all structures from rising RT plumes, the apparent boxy profile suggests 
that the rise of such plumes may be significantly limited. 
 
Given   implications of the [\ion{Fe}{2}] line profile, it is worthwhile to test
to see if the profile is present in other spectral features.  
Unfortunately, the S/N in the 1.257~\micron\
feature is not sufficient to place much of a constraint on the line profile, though it is not inconsistent
with the 1.644~\micron\ feature.
In the optical spectrum, the features appear more peaked, seemingly in contradiction to the IR
feature.  However, the features in the 4000--5000 \AA\ are composed of blends of large numbers of 
forbidden lines from iron-group elements.
  Thus the profiles of these features are rather weakly linked to
the kinematic distribution of radioactive material and,
in our model at day 300, do not allow to analyze its boxiness \footnote{Note that the profile will change
with time and we did not study the time-evolution or a wide range of models}.

The only relatively clean [Fe II] feature (7155 \AA )  seems to contradict the observed 1.644~\micron\
profile, showing what
seems to be a definite peak at 7200 \AA (Fig~\ref{model_optical}).   Indeed such a profile seems more suggestive of our mixed
model (red) than the reference model with the central hole in the \nick\ distribution (blue).
However, neither interpretation explains the extended red wing of the observed emission.
Instead, both the peaked emission at 7200~\AA\ and the extended red wing extending to about 7550 \AA\ 
can be well reproduced by two blended flat-topped profiles, consistent with the observed NIR line, produced 
by [\ion{Fe}{2}] 7155~\AA\ and blends of weaker [\ion{Fe}{2}] 7438~\AA\ and  [\ion{Ca}{2}] 7291 \& 7324 \AA\ which have an
intrinsic emission  ratio of 3:2.
 Note that the relative contribution to the red feature depends on the Ca abundance of the specific model.
Some support in favor of a significant CaII contribution may be taken from the observations because a sufficiently
strong Fe feature at 7300~\AA\ would produce a secondary, red  feature which has not been observed in 
SN~2003du (see Fig. \ref{model_optical}).
The combined blend explains the red wing of the 'plateau' as the edge of the flat-topped
[\ion{Ca}{2}] line and, at the same time produced the steep red wing and flatter blue wing
of the ``[\ion{Fe}{2}]'' line through  blends.
Nevertheless, such ambiguity in the optical demonstrates the need of Near-IR observations at late time.

\subsection{Deflagration scenario}
\begin{figure*}
\includegraphics[width=13.7cm,angle=0,clip=true,trim=-60bp 0bp 0bp 00bp]{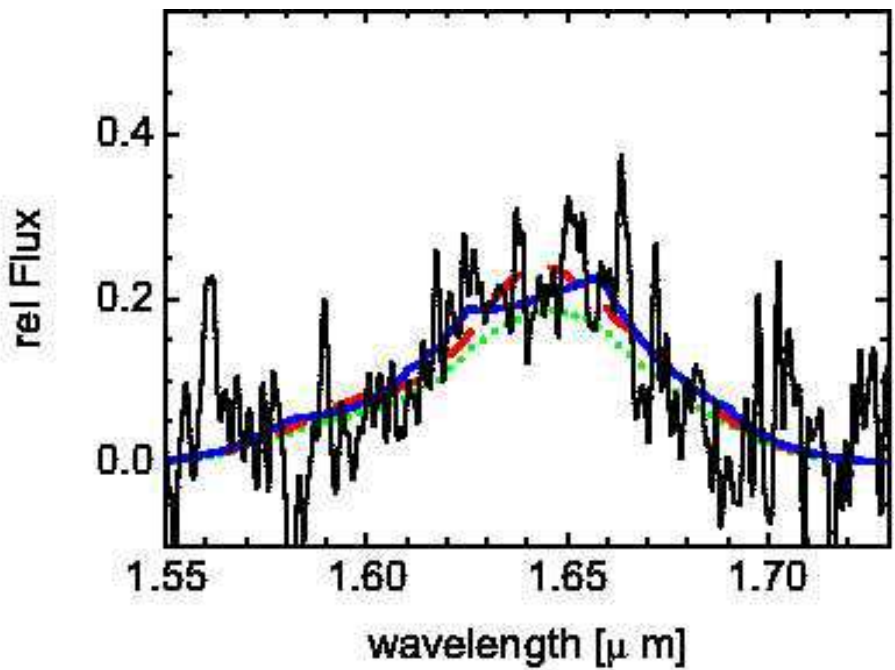}
\caption{A flat-topped profile can be produced by any $M_{ch} $ model 
without emission for velocities $\leq 3000$ \kms.  We show a comparison of the observed [\ion{Fe}{2}] 1.644 \micron  ~
feature and a calculated line profile based on the deflagration model W7 with some variations \citet{nomoto84}.
We show the results for the original, unmixed model (dotted),  with mixing the layers below 8000\kms\ (dashed), and assuming a hole
in the emission for layers with  velocities less than 3000 \kms\ (solid). The
 fluxes are  normalized to the peak of the unmixed model
without adjusting the profiles for the other model to the observed data.
}
\label{w7}
\end{figure*}

 Based on the spherical deflagration model W7, a similar analysis of the data has been carried out by Maeda using
the code of Ruiz-Lapuente \& Lucy (1992). In Fig. \ref{w7},
 we show a comparison of the observed [\ion{Fe}{2}] 1.644 $\micron $
feature and a calculated line profile based on the deflagration models W7 with some variations \citet{nomoto84}.
We show the results for the original, unmixed model (dotted),  with mixing the layers below 8000\kms\ (dashed), and assuming a hole
in the emission for layers with  velocities less than 3000 \kms\ (solid) with the flux normalized to the peak of the unmixed model
without adjusting the profiles for the other model to the observed data.
 At day 300 and the resulting profile is not entirely flat topped due to two effects: first in
 W7, \nick\ can be found down to about 2200 to 2300  \kms\ and, second, this calculation still exhibits significant 
  energy deposition by $\gamma $-rays near the center.
 In part, the central $\gamma$-ray deposition is caused by the assumption that each scattered  results in full thermalization.
Apparently, we require a slightly larger hole of 3000 \kms\ to create a sufficiently broad, flat topped feature.

\section{Alternate Interpretations}
Our suggestion that the flat-topped profiles of SN~2003du indicate a concentration of electron-capture
isotopes in the central region with no or little mixing in of \nick, is not the only possible
interpretation. 
%
For example, in the `infrared catastrophe' picture \citep{a80a},
a lack of emission at certain expansion velocities can be caused by a 
redistribution of energy from the optical/near IR to the far IR when the density drops below
a certain limit (about 100 days after the explosion). However, the brightness of SN~2003du is consistent with
the predictions from early times rather than showing a rapid drop. Furthermore, the outer, low-density, 
high-velocity layers should exhibit the effect first, and we see no
evolution of the line profiles between Feb~27 and Apr~2nd, 2004.  Moreover, the optical
spectrum looks typical for late-time spectra. Thus, as none of the other predicted signatures of an IR
catastrophe can be identified in the observations, we regard this alternative explanation as unlikely.

A difference in the ionization structure of iron from the one assumed here could be another
explanation since the observed line profile depends on the
\ion{Fe}{2}\ distribution and, in principle, a shift in the ionization state to either Fe I or III
may produce flat-topped Fe~II profiles.
 In principle, a shift to Fe~III may show up in the region around 4700 \AA. However, in practice,
such analysis is severely hampered by strong blending in this region which produce very wide features
equivalent to $\approx 20,000$ \kms\  compared to $3,000 km~s^{-1}$ considered here.
 Deconvolving such a feature from this blend depends sensitively on details of
the ionization and excitation structure. In our observations, we do not see evidence for a strong narrow
emission component by [Fe~III] in this region though, empirically, some contribution cannot be ruled out.


Detailed calculations for the ionization structures at about day 300 
have been published for W7, a model with a chemical structure similar 
to our delayed detonation model (RL95; LJS97). In these models 
the ionization balance shifts from \ion{Fe}{2} to \ion{Fe}{3} from the inner to the outer regions 
because the recombination rate and charge exchange
reactions decrease quadratically with the density in the region with radioactive isotopes.
As a consequence, higher ionization stages are seen at larger radii, although the   
\ion{Fe}{2}/\ion{Fe}{3} ratio is rather smoothly decreasing with density, without 
sudden changes in the ionization balance.
Thus, the lack of \ion{Fe}{2} emission from the central layers cannot
be attributed to a change in the balance towards \ion{Fe}{3}.
Furthermore, in the radioactive region, ionization by (local) positron deposition keeps the
\ion{Fe}{2}/\ion{Fe}{1} ratio high. In presence of mixing, we cannot expect a 'hole' in the
\ion{Fe}{2} distribution which leads us back to our basic scenario of a central hole in the $^{56}$Ni distribution.

 The necessity of a hole of 3000 \kms\ in the \nick\ distribution is not unique to these specific models we have
examined here. Any $M_{ch}$ model will require a similar hole to be consistent with the observation.
The size of the central region with elements produced by electron capture may, in principle,
provide a measure for both the central densities and possible mixing. Currently however, uncertainties
in the treatment of the deflagration front and the electron capture rate do not allow us to discriminate
between specific models \citep{bra00}.

\begin{figure*}
\includegraphics[width=14.7cm,angle=0,clip=true]{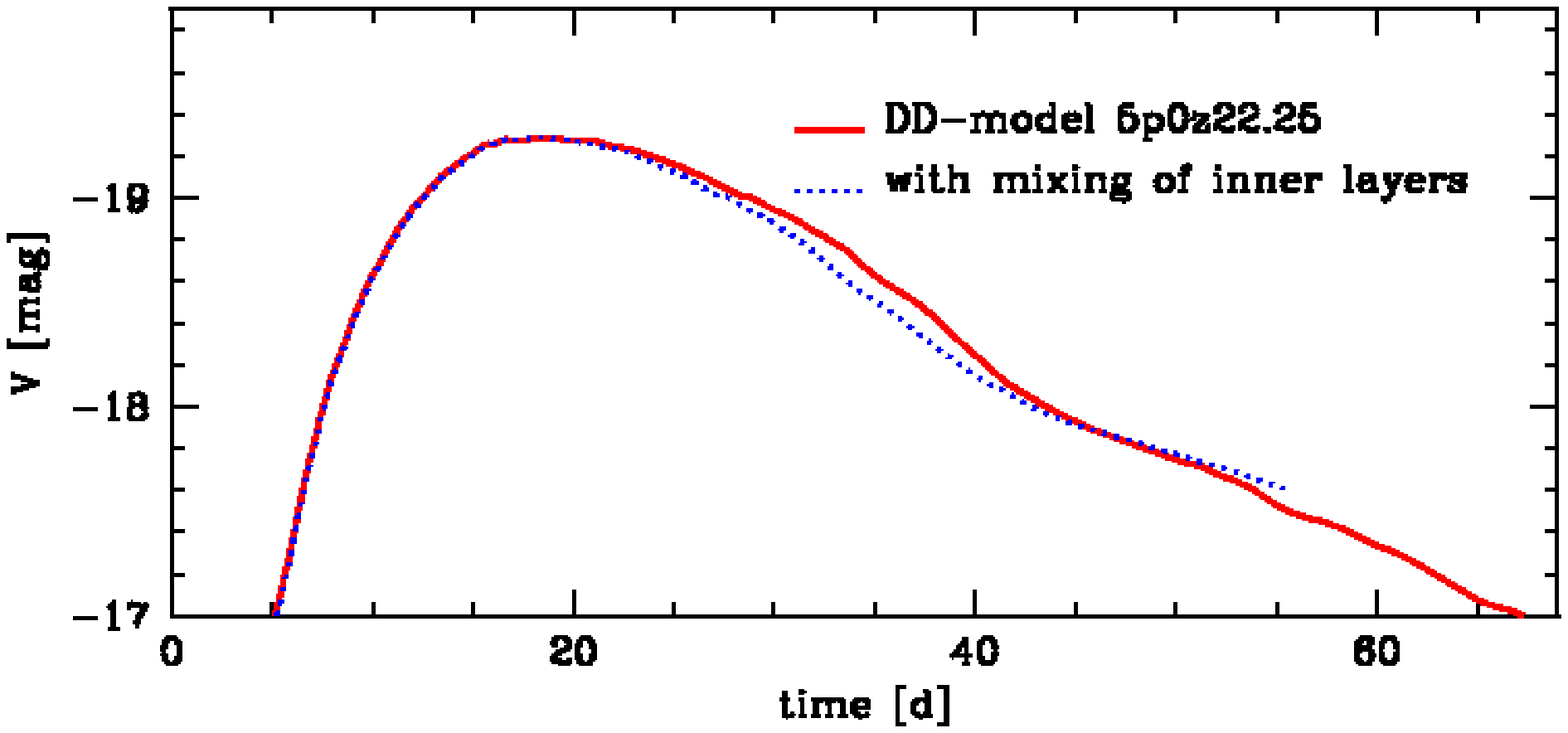}\\
\includegraphics[width=14.7cm,angle=0,clip=true]{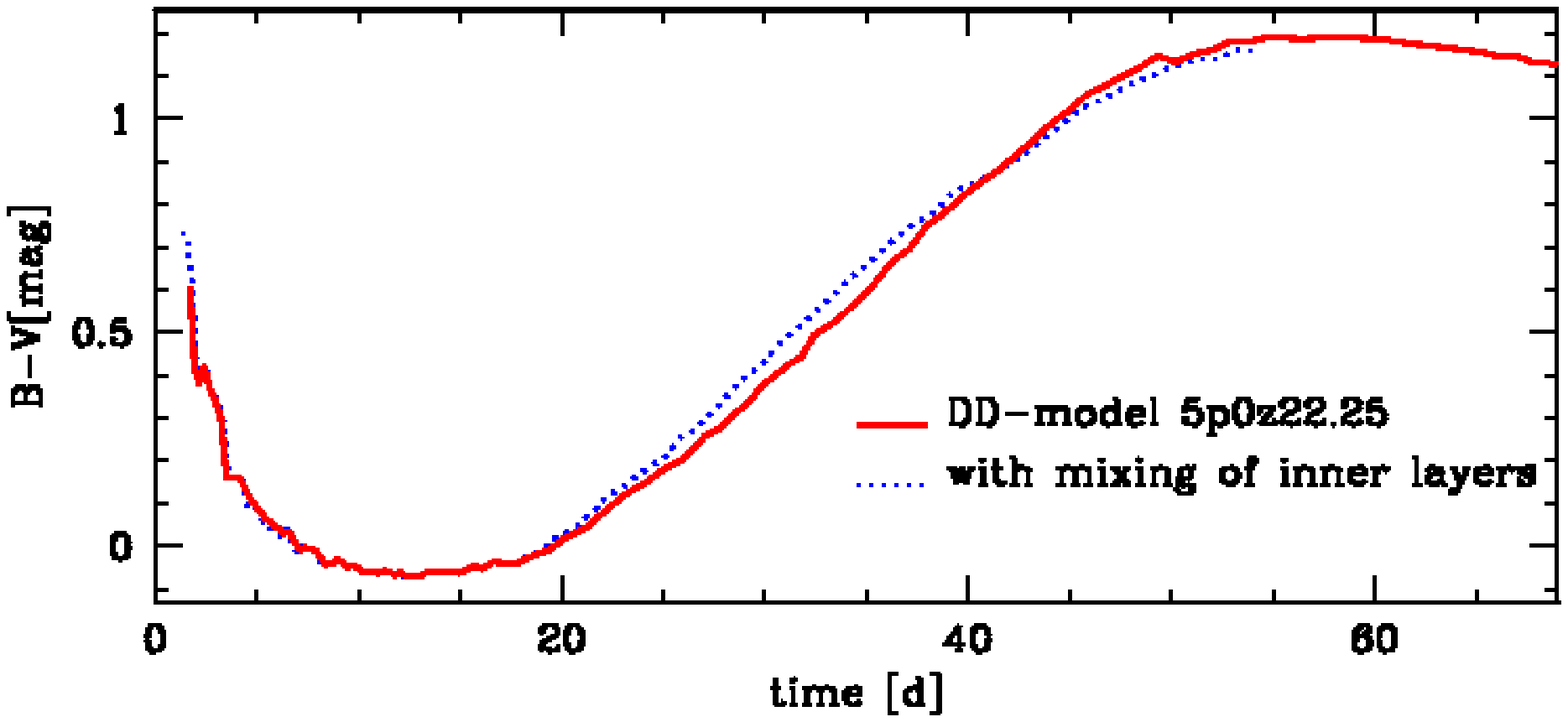}
\caption{Visual light curves and $(B-V)$ for the same delayed detonation model but with
and without mixing of the inner layers (expansion velocities $\leq 8000$\kms).
Our LC calculations take into account detailed hydrodynamics, nuclear networks,
radiation transport for low and high energy photons, opacities and include solvers to calculate 
the atomic level populations \citep{hoeflich03a}.}
\label{lc}
\end{figure*}

\section{Discussion}

\subsection{SN~Ia Explosion Physics}
Although the observed profile can be reproduced by spherical models of massive WDs,
our results pose a serious challenge to our understanding of the detailed physics of 
the explosion.  All current generation 3-D models predict deflagration fronts with 
properties dominated by slowly-growing large-scale Rayleigh-Taylor instabilities which 
allow the products of high-density nuclear burning to rise in plumes and mix with \nick. 
However, our results suggest that the early phase of the explosion, which leads to the
early expansion of the WD, is not dominated by RT instabilities (at least not in the case of SN~2003du),
though there were similar suggestions from the analysis of SN~1999by a very subluminous SNe~Ia.

  Perhaps the early phases of nuclear burning are dominated by pre-existing velocity fields 
produced prior to the thermonuclear runaway \citep{hs02}. Alternatively, this phase might be dominated by the rise
of a single large bubble of burned material which penetrates the WD surface and triggers
a detonation that propagates inward (Plewa et al. 2004).
We may speculate that variations in the central density of the WD may be responsible for the 
appearance of profiles like that apparently seen in SN~2003du or, even more radically, 
that the pre-expansion of the WD prior to the detonation phase is not produced during the 
explosive phase of burning but rather during the pre-condoning phase followed
directly by an explosive detonation phase.

\subsection{Comparison With Other SNe~Ia}
It is useful to put our findings into context and examine the signatures of mixing the very inner layers
in other SNe~Ia.  As mentioned above, the peculiar SN~1991T is the only other SNe~Ia for which late-time 
($> 200$~d) NIR 
line profiles have been published. In contrast to our observations of SN~2003du, the line profiles in SN~1991T
were triangular-shaped, strongly indicating mixing of the inner layers.  Late-time observations of the ''normal-bright"
  SN~1998bu  (Meikle 2004, Spyromilio et al. 2004) also seem to show a more centrally peaked profile in the NIR
[\ion{Fe}{2}] lines \footnote{However, e.g. the optical spectrum of SN~2003cg around 400 days after maximum light
shows a very boxy shape to the [\ion{Fe}{2}]/[\ion{Ca}{2}] emission feature near 7200 \AA\ (Elias de la Rosa et al., 2004)}.

Note that in the scenario we suggest here, even in a SN with very little mixing, the flat-topped
profile will be a transient phenomenon.  At earlier times, the central regions of the ejecta will
still be optically thick to $\gamma$-rays, and will therefore show significant low-velocity emission.
As the inner layers become optically thin to $\gamma$-rays, the central peak would be expected to fade,
leaving a boxy profile like the one discussed here, if the magnetic field in the ejecta is strong enough
to prevent the free streaming of fast electrons from the radioactive ejecta into the center region.  Eventually
the expansion of the ejecta would dilute the magnetic field sufficiently to allow the penetration of fast
electrons.  The energy deposition would no longer be local, and the central peak would light up again.  
Thus, this scenario would predict an evolution where the [\ion{Fe}{2}] features start centrally peaked, 
proceed through a phase of boxy emission, and then return to a peaked emission phase (unless the magnetic
field in the WD progenitor is too weak to cause the local trapping of fast electrons, in which case the 
profile would always remain peaked.)  Indeed, if such an evolution would be observed in the late-time
spectra of a Type~Ia SN, the epoch of the second transition to a peaked profile could, in principle, 
be used to infer the strength of the magnetic field in the central region of the WD progenitor.

The very subluminous SN~1999by 
presented a different constraint on the kinematics of the \nick-rich ejecta. 
In subluminous SNe~Ia, little \nick\ is produced during the detonation and the Si-rich region extends down to
expansion velocities of $\approx 5000$~\kms. As a consequence, the chemical signatures of the deflagration
phase will survive the subsequent detonation and has been expected to show up as iron-rich plumes embedded
in a Si/S rich region. However, strong mixing can be excluded based on the spectral evolution of SN~1999by at 
early times \citep{howell01,h02}. 

\subsection{Implications for SN~Ia Cosmology}
Our results not only pose a challenge to our present understanding of the physics of
the explosion but also raise questions of whether SN~2003du is an exception or the norm, and how this
will affect our understanding of the diversity of SNe~Ia. The implications for the use of supernovae in high 
precision cosmology may be significant as it requires a 2 \% accuracy in distance \citep{os01,wa01}.
 In Fig. \ref{lc}, light curves are shown for two explosion models 
(Fig.~\ref{struct}) with and without full mixing of layers with expansion velocities $\leq $
$8000$ \kms\ (guided by 3-D calculations \citep{k01}), but otherwise identical. 
Such mixing will bring some \nick\ to the central, high density region and reduces
the \nick\ abundance at other layers. Quantitatively, the results will depend
details of the mixing, and the differences are of the same order as the numerical accuracy of our calculation. Thus, the 
LCs are shown here to demonstrate the qualitative effects and the approximate size of the effects.

The difference can be can be understood in terms of energy deposition by
the $\gamma$-rays and the diffusion time scales of low energy photons in the SN envelope. The latter
strongly decreases with time {\it t} (roughly $ \propto t^{-2}$ \citealt{h95}) and, at maximum
light, it is comparable to the expansion time scale.
Up until about 2 to 3 days before maximum light, the luminosity is mostly determined by the
outer part of the \nick\ layer and, consequently, the early light curves are virtually identical.
With time, deeper layers start to contribute. Because mixing effectively reduces the \nick\ 
which contributes to the luminosity prior to about day 30, the LC of the mixed model declines faster 
and is redder.  A competing, opposite effect is related to the energy deposition by $\gamma$-rays or,
more precisely, the escape probability for high energy photons. Increasingly, $\gamma$-rays produced
in the outer layers of the $^{56}$Ni distribution can escape and, as a consequence, are lost for
heating the envelope.  In SNe~Ia, the escape probability for $\gamma$-rays increases strongly with time from
about 10 to 15\% at maximum light to about 60\% at about day 50 (H\"oflich et al. 1992).
Mixing of \nick\ increases the trapping of $\gamma$-rays because the inner layers are optically
thick. As a consequence, the luminosity of the mixed model is higher and the temperature slightly higher.
In conclusion, mixing processes will affect the brightness decline relation and the peak to tail ratio
on the 0.1 mag level. Depending on the amount of mixing and its dependence on the progenitor,
this may cause small but systematic evolutionary effects with redshift.  
\section{Summary}

We have presented late-time near-infrared and optical spectra of the  `normal-bright'
Type Ia supernova 2003du. We show
that late-time infrared spectra provide a probe
of the distribution of radioactive \nick\ in the ejecta. The strong, non-blended [Fe~II] 
emission line at 1.644 \micron\ is particularly well suited for such work, and in SN~2003du it
appears to exhibit a box-like profile, whereas the optical [Fe~II] features (including the strong, 
single [Fe~II] line near 7200~\AA) appear more peaked like those previously seen in other SNe~Ia 
at comparable epochs.

Based on detailed models for exploding Chandrasekhar-mass white dwarfs, we have
presented a quantitative analysis of the observations, including optical depth effects
(although these are rather small). Using our detonation models, we have shown that the
line profile of the 1.644 \micron\ line is consistent with spherical explosion models
which produce a central region electron-capture elements up to $\approx 2500$--3000~\kms\
as a consequence of burning under high density, and an extended region of radioactive \nick\ 
up to $\approx 9,000$--10,000 \kms.
We have shown that the observed IR line profile is inconsistent with mixing of the region of
neutron-rich isotopes with the \nick-rich layers.
The 1.644 \micron\ line is offset about 500 \kms\ relative to the host galaxy which might indicate
a slight shift of the distribution of radioactive elements with respect to the envelope.

The observed peaked profile of the optical [Fe~II] feature at 7200~\AA  
and its extended wing can be well reproduced by a blend of [Fe~II] 7155 \& 7172~\AA\ and weaker lines  around
7388 \& 7452 ~\AA\  ,  and
[CaII] 7291/7324~\AA, with line profiles consistent with the 1.644 \micron\ feature.
Such ambiguity in the interpretation of the optical lines demonstrates the usefulness of 
near-IR observations beyond day 250--300, at phases when $\gamma $-ray
deposition becomes unimportant for the ionization balance of iron group elements.

Although a hole in the \nick\ seems to be a necessary condition to
produce a flat-topped line profile, we also need to require sufficiently local trapping 
of fast electrons, as could be produced by weak magnetic fields.  Higher signal-to-noise 
observations of other normal-bright SNe~Ia, coupled with 
detailed ionization models are certainly needed and will provide insight into the diversity
(and causes thereof) in SNe~Ia, and also into possible sources of evolutionary effects at higher redshift.

While the flattened [Fe~II] 1.644 $\mu m$ profile is consistent with our 1-D models,
it presents a challenge to our understanding of the physics in the earliest phase of the explosion, since all current
3-D models of the deflagration phase predict significant mixing in the innermost regions. This
result for SN~2003du, while differing significantly from the late-time observations of the
unusual SN~1991T, places constraints on the mixing of \nick\ during the deflagration which are
similar to constraints placed on the mixing in the extremely sub-luminous SN~1999by.  
We now have examples of SNe~Ia in which mixing may be suggested (SN~1991T, SN~1998bu) and in which it is
excluded, or at least very much limited (SN~1999by and SN~2003du).  Understanding these phenomena 
will have significant implications both for the physics of SN~Ia explosions, and for their use
as cosmological probes.

 Finally, we also have to address the limitations of this study which  are subject to ongoing projects.
 Programs at SUBARU will increase the number  of SNe~Ia with late-time IR-observations and to study the
spectral evolution with time.  Detailed calculations are under way for the ionization structure which include
magnetic fields with the goal to study their influence on the spectra and the spectral evolution.

\acknowledgements
We wish to thank the staff at Subaru and McDonald observatory for their excellent support, especially 
K. Aoki and D. Doss who were instrumental in getting these observations completed.  We also thank the 
Subaru Director's office for giving us a second chance to try and observe SN~2003du in April.
 The light curve calculations have been performed on the remaining 11 nodes of the Beowulf cluster at the
Department of   Astronomy at the University of Texas which has been financed by the John W. Cox Fund in 1999, and
maintained by the NASA grant NAG5-7937 to PAH.  This work was supported by NSF grant AST0307312 to PAH,
NSF AST-9876703 to RSF, and the grant-in-Aid for Scientific Research (15204010, 16042201, 16540229) of MEXT, Japan, to KN.


\begin{thebibliography}{} 
\bibitem[Axelrod(1980a)]{a80a} Axelrod, T.~S.\ 1980a, Texas Workshop on Type I Supernovae, 80
\bibitem[Axelrod(1980b)]{a80b} Axelrod, T.~S.\ 1980b, Ph.D.~Thesis
\bibitem[Bowers et al.(1997)]{b97} Bowers, E.~J.~C., Meikle, W.~P.~S., Geballe, T.~R., Walton, N.~A., Pinto, P.~A., Dhillon, V.~S., Howell, S.~B., \& Harrop-Allin, M.~K.\ 1997, \mnras, 290, 663
\bibitem[Brachwitz et al.(2000)]{bra00} Brachwitz, F., et al.\ 2000, \apj, 536, 934
\bibitem[Branch(1999)]{branch99} Branch, D. 1999, \araa, 36, 17
\bibitem[Elias de la Rosa et al.(2004)]{Elias04} Elias de la Rosa N., Turatto M., Cappellaro E., Stanishev V., Kotak R., Pignata G. 2004, 
in: Supernovae as Cosmic Lighthouses,  eds. Turatto et al., PASP, in preparation
\bibitem[Fisher et al.(1995)]{fisher95} Fisher, A., Branch, D., H\"oflich, P., Khokhlov, A. 1995, ApJ 447, 73
\bibitem[Fransson, Lunquist \& Chevalier(1996)]{fransson96} Fransson, C., Lundqvist, P., Chevalier,   	R. A. 1996, \apj, 461, 993
\bibitem[Gamezo et al.(2003)]{g03} Gamezo, V.~N., Khokhlov, A.~M., Oran, E.~S., Chtchelkanova, A.~Y., \& Rosenberg, R.~O.\ 2003,   Science, 299, 77
\bibitem[Gamezo, Khokhlov, \& Oran(2002)]{g02} Gamezo, V.~N., Khokhlov, A.~M., \& Oran, E.~S.\ 2002, Bulletin of the American Astronomical Society, 34, 663
\bibitem[Gerardy et al.(2004)]{gerardy04} Gerardy, C. L., H\"oflich, P., Fesen, R. A., Marion, G. H., Nomoto, K., Quimby, R., Schaefer, B. E., Wang, L., \& Wheeler, J. C. 2004, \apj, in press (astro-ph/0309639)
\bibitem[Gerardy \& Fesen(2001)]{gf01} Gerardy, C.~L.~\& Fesen, R.~A.\ 2001, \aj, 121, 2781
\bibitem[H\"oflich(1995)]{h95} H\"oflich, P. 1995, \apj, 443, 89
\bibitem[H\"oflich \& Khokhlov(1996)]{hk96}  H\"oflich, P., Khokhlov, A. 1996, \apj, 457, 500
\bibitem[H\"oflich et al.(1996)]{h96} H\"oflich, P., Khokhlov, A., Wheeler, J. C., Phillips, M. M., 	Suntzeff, N. B., \& Hamuy, M. 1996, \apj, 472, L81
\bibitem[H\"oflich, Wheeler, \& Thielemann(1998)]{hwt98}Hoeflich, P., Wheeler, J.~C., \& Thielemann, F.~K.\ 1998, \apj, 495, 617
\bibitem[H\"oflich et al.(2002)]{h02} H\"oflich, P., Gerardy, C., Fesen, R., \& Sakai, S. 2002,   	\apj, 568, 791
\bibitem[H\"oflich \& Stein(2002)]{hs02} H\"oflich, P., Stein, J. 2002,   	\apj, 568, 771
\bibitem[H\"oflich et al.(2003)]{hetal03} H\"oflich, P., Gerardy, C., Linder, E., \& Marion, H.  	2003, in: Stellar Candles, 
eds. Gieren et al., Lecture Notes in Physics 635, Springer Press, p. 203 \& astro-ph/0301334
\bibitem[H{\" o}flich(2003)]{hoeflich03a} H{\" o}flich, P. 2003, ASP Conf.~Ser.~288: Stellar Atmosphere Modeling, 185
\bibitem[H{\" o}flich(2003)]{hoeflich03b} H{\" o}flich, P. 2003, ASP Conf.~Ser.~288: Stellar Atmosphere Modeling, 371
\bibitem[Howell et al.(2001)]{howell01} Howell, A., H\"oflich, P., Wang, L., Wheeler, J. C. 2001,  	\apj, 556, 302
\bibitem[Iwamuro et al.(2001)]{iwamuro01} Iwamuro, F., Motohara, K., Maihara, T., Hata, R. \& Harashima, T. 2001, \pasj, 53, 355
\bibitem[Khokhlov(2001)]{k01}  Khokhlov, A. 2001, astro-ph/0008463
\bibitem[Khokhlov(1995)]{k95} Khokhlov, A.~M.\ 1995, \apj, 449, 695
\bibitem[Khokhlov(1991)]{k91} Khokhlov, A. 1991, \apj, 245 114
\bibitem[Kotak \& Meikle(2003)]{kotak03} Kotak, R. \& Meikle, W. P. S. 2003,   \iaucirc\ 8122
\bibitem[Kurucz(1993)]{kurucz93} Kurucz, R. L. 1993, Atomic Data for Opacity Calculations,  	Cambridge/Center for Astrophysics, CD 1
 \bibitem[Kurucz(1994)]{kuruzc94} Kurucz, R. L. 1994, Solar Abundance Model      Atmospheres for 0,1,2,4,8 km/s, Kurucz CD-ROM No.\ 19
\bibitem[Leibundgut(1995)]{leib95} Leibundgut B. 1995, Proceedings of the 34th Herstmonceux Conference, eds. Clegg et al., Cambridge University Press 1994, p.100
\bibitem[Lentz et al.(2001)]{lentz01} Lentz, E. J., Baron, E., Branch, D., Hauschildt, P., Nugent,  	P. E. 2001, \apj, 530, 966L
\bibitem[Lisewski et al.(2000)]{l00} Lisewski, A.~M., Hillebrandt, W., Woosley, S.~E., Niemeyer, J.~C., \& Kerstein, A.~R.\ 2000, \apj, 537, 405
\bibitem[Liu, Jeffery, \& Schultz(1997)]{l97} Liu, W., Jeffery, D.~J., \& Schultz, D.~R.\ 1997, \apjl, 483, L107 (LJS97)
\bibitem[Livne \& Arnett(1993)]{liv93} Livne E., Arnett D. 1993, ApJ 415, L107
\bibitem[Livne (1999)]{liv99} Livne E. 1999, ApJ 527, L97
\bibitem[Maeda et al(2003)]{maeda03} Maeda, K., Mazzali, P., Deng, J., Nomoto, K., Yoshhi, Y., 	Tomika, H., Kobayashi, Y. 2003, \apj, 593, 22
\bibitem[Massey et al(1988)]{massey88} Massey, P., Strobel, K., Barnes, J. V., \& Anderson, E. 1988, \apj, 328, 315
\bibitem[Meikle (2004]{m04} Meikle P., 2004, in: Supernovae as Cosmic Lighthouses,  eds. Turatto et al., PASP, in preparation
\bibitem[Meyerott(1980)]{m80} Meyerott, R.~E.\ 1980, \apj, 239, 257 
\bibitem[Motohara et al.(2002)]{motohara02} Motohara, K., et al.\ 2002, \pasj, 54, 315
\bibitem[Niemeyer \& Woosley(1997)]{n97} Niemeyer, J.~C.~\&  Woosley, S.~E.\ 1997, \apj, 475, 740
\bibitem[Nomoto(1982)]{nomoto82} Nomoto, K. 1982 \apj, 253, 798
\bibitem[Nomoto, Thielemann, \& Yokoi(1984)]{nomoto84} Nomoto, K., Thielemann, F. -K., \& Yokoi, K.   	1984, \apj, 286, 644
\bibitem[Nomoto et al.(2000)]{nomoto00} Nomoto, K., Umeda, H., Hachisu, I., Kato, M., Kobayashi, C.,  \& Tsujimoto, T., 2000, in Type Ia Supernovae: Theory and Cosmology, ed.\ J. Truran, J. Niemeyer  (Cambridge: Cambridge University Press), 63
\bibitem[Nomoto et al.(2003)]{nomoto03} Nomoto, K., Uenishi, T., Kobayashi, C., Umeda, H., Ohkubo,  T., Hachisu, I., \& Kato, M. 2003, in From Twilight to Highlight: The Physics of Supernovae, eds. W. Hillebrandt \& B. Leibundgut, ESO Astrophysics Symposia (Berlin: Springer), 115  (astro-ph/0308138)
\bibitem[Nussbaumer \& Storey(1980)]{ns80} Nussbaumer, H., \& Storey, P. J.\ 1980, \aap, 89, 308
\bibitem[Nussbaumer \& Storey(1988a)]{ns88a} Nussbaumer, H.~\& Storey, P.~J.\ 1988, \aap, 193, 327
\bibitem[Nussbaumer \& Storey(1988b)]{ns88b} Nussbaumer, H.~\& Storey, P.~J.\ 1988, \aap, 200, L25
\bibitem[Oliva, Moorwood, \& Danziger(1989)]{omd89} Oliva, E., Moorwood, A. F. M., \& Danziger, I. J.\ 1989,
\aap, 214, 307
\bibitem[Oliva, Moorwood, \& Danziger(1990)]{omd90} Oliva, E., Moorwood, A. F. M., \& Danziger, I. J.\ 1990,
\aap, 240, 453
\bibitem[Ostriker \& Steinhardt(2001)]{os01} Ostriker P., Steinhardt P.J. 2001, Sci.Am., 284, 47
\bibitem[Phillips(1993)]{p93} Phillips, M.M., 1993, \apj, 413, 105
\bibitem[Reinecke, Hillebrandt, \& Niemeyer(2002)]{r02} Reinecke, M., Hillebrandt, W., \& Niemeyer, J.~C.\ 2002, \aap, 391, 1167
\bibitem[Reinecke, Hillebrandt, \& Niemeyer(1999)]{r99}  Reinecke, M., Hillebrandt, W., \& Niemeyer, J.~C.\ 1999, \aap, 347, 739
\bibitem[Ruiz-Lapuente et al.(1995)]{ruiz95} Ruiz-Lapuente, P., Kirshner, R.~P., Phillips, M.~M., Challis, P.~M., Schmidt, B.~P.,  Filippenko, A.~V., \& Wheeler, J.~C.\ 1995, \apj, 439, 60 (RL95)
\bibitem[Ruiz-Lapuente \& Lucy(1992)]{ruiz92} Ruiz-Lapuente, P., Lucy, L. 1992, ApJ 400, 127
\bibitem[Schneider et al.(1992)]{schneider92} Schneider, S. E., Thuan, T. X., Magnum, J. G., \& Miller, J.
1992, \apjs, 81, 5
\bibitem[Schwartz \& Holvorcem(2003)]{schwartz03} Schwartz, M. \& Holvorcem, P.  R. 2003, \iaucirc\ 8121
\bibitem[Spyromilio et al.(2004)]{s04} Spyromilio, Sollerman J., Leibundgut B., Fransson C., Cuby J.-C. 2004, A\&A, in press
\bibitem[Spyromilio, Meikle, Allen, \& Graham(1992)]{s92} Spyromilio, J., Meikle, W.~P.~S., Allen, D.~A., \& Graham, J.~R.\ 1992, \mnras, 258, 53P 
\bibitem[Tody(1986)]{tody86} Tody, D. 1986, Proc.\ SPIE, 627, 733
\bibitem[van Dokkum(2001)]{vandokkum01} van Dokkum, P. G. 2001, \pasp, 113, 1420
\bibitem[Weller J. \& Albrecht(2001)]{wa01} Weller J., Albrecht A. 2001, Physical Review Letters, 86, 1939
\bibitem[Wheeler et al.(1998)]{wheeler98} Wheeler, J. C., H\"oflich, P., Harkness, R. P., Spyromilio 	J. 1998, \apj, 496, 908
\bibitem[Woosley \& Weaver(1994)]{ww94} Woosley, S. E., \& Weaver, T. A. 1994, \apj, 423, 371
\bibitem[Yamaoka et al.(1992)]{yamaoka92} Yamaoka, H., Nomoto, K., Shigeyama, T., \& Thielemann, 	F. 1992, \apj, 393, 55
\end{thebibliography}
\end{document}